\documentclass[twocolumn,prl,aps,showpacs,superscriptaddress]{revtex4}
\usepackage{amsmath,amssymb,amsfonts,bbm}

\newcommand{\ket}[1]{| #1 \rangle}

\newcommand{\ketbra}[2]{  | #1 \rangle \langle #2 |}
\newcommand{\tr}[1]{\mbox{Tr}( #1)}
\renewcommand{\[}{\begin{equation}}
\renewcommand{\]}{\end{equation}}
\newcommand{\pt}[1]{ #1^{\mbox{\tiny \textup{pt}}}    }
\newcommand{\id}{\ensuremath{\mathbbm 1}}

\begin{document}

\title{Multipartite Entanglement Criterion from Uncertainty Relations}

\author{J. Gillet}
\affiliation{Institut de Physique Nucl\'eaire, Atomique et de Spectroscopie, Universit\'e de Li\`ege, 4000 Li\`ege, Belgium}
\author{T. Bastin}
\affiliation{Institut de Physique Nucl\'eaire, Atomique et de Spectroscopie, Universit\'e de Li\`ege, 4000 Li\`ege, Belgium}
\author{G. S. Agarwal}
\affiliation{Department of Physics, Oklahoma State University, Stillwater, OK 74078-3072, USA}

\date{\today}

\begin{abstract}
We formulate an entanglement criterion using Peres-Horodecki positive partial transpose
operations combined with the Schr\"odinger-Robertson uncertainty
relation. We show that any pure entangled bipartite and tripartite state can be detected
by experimentally measuring mean values and variances of specific observables.
Those observables must satisfy a specific condition in order to be used, and
we show their general form in the $2\times 2$ (two qubits) dimension case. The criterion is
applied on a variety of physical systems including bipartite and multipartite mixed states
and reveals itself to be stronger than the Bell inequalities and other criteria. The criterion also work
on continuous variable cat states and angular momentum states of the radiation field.

\end{abstract}

\maketitle

In the past few years, many criteria detecting entanglement in
bipartite and multipartite systems have been
developed~\cite{Min-Bra-Pat}. The Peres-Horodecki
positive partial transpose (PPT) criterion~\cite{Per96} has played a crucial role in the field and
provides, in some cases, necessary and sufficient conditions to entanglement. That criteria
is formulated in terms of the density operator and any practical application involves state tomography.
Other criteria have been proposed so they could be tested
experimentally in a direct manner, as the Bell inequalities~\cite{Bell, Dur01} or the entanglement witnesses~\cite{Guh02}. More
recently, criteria based on variance measurements have been
studied for continuous and discrete variable systems~\cite{contin,Hil06,discr,both,Guh04, Aga05, Nha06, Nha07,Son08}.

In~\cite{Aga05} the Heisenberg
relation has been used along with the partial transpose operation to obtain a
criterion detecting entanglement condition in bipartite non-gaussian states. That idea was generalized in~\cite{Nha07,Son08} with use of the Schr\"odinger-Robertson
relation instead of the Heisenberg inequality.  In this paper, we generalize completely those concepts and prove that the Schr\"odinger-Robertson type inequality is
able to detect entanglement in any pure state of bipartite and tripartite systems.
Experimentally, it can be realized by measuring mean values and variances of different observables;
however we show that all observables are not suitable
and we yield the general condition they must
satisfy to be eligible. For $2\times 2$ systems, we explicitly give
their general form. The inequality has a wide application range~:
qubits, angular momentum states of harmonic oscillators, cat states, etc.
For the mixed state case, the inequality detects
entanglement of bipartite Werner states better than the Bell inequalities~\cite{Bell}
and also leads to a good characterization of multipartite Werner states.

For any observables $A, B$ and any density operator $\rho$, the
Schr\"odinger-Robertson uncertainty relation reads~\cite{Sch-Rob}
\[  (\Delta A)^2 (\Delta B)^2 \ge \frac{1}{4} | \langle [ A,B] \rangle |^2 +  \frac{1}{4} | \langle \{ A,B \} \rangle - 2 \langle A \rangle \langle B \rangle |^2 \label{eq-1}, \]
where $\langle A \rangle = \tr{\rho A}$ is  the mean value of $A$,
$(\Delta A)^2=\tr{\rho A^2}-\tr{\rho A}^2$ its variance, $[A,B]$ the
commutator and $\{ A,B \} \equiv AB+BA$ the anticommutator of $A$
and $B$. The Heisenberg uncertainty relation is obtained if the last
term is not considered, which gives a weaker inequality.

The PPT criterion~\cite{Per96} is a sufficient condition for
entanglement, saying that if a bipartite state $\rho$ is separable
it can be written as $\rho = \sum_i p_i \, \rho^1_i \otimes\rho^2_i$ with usual notations and its partial transpose $\pt \rho
\equiv \sum_i p_i \, \widetilde \rho^1_i \otimes\rho^2_i$ must be positive. In the case of multipartite systems, we also consider the partial transposition of the first subsystem.

The partial transpose $\pt\rho$ of a bipartite separable density
operator $\rho$ must be positive, which implies it does describe some
physical state and must therefore obey the
Schr\"odinger-Robertson
uncertainty relation for any observables $A$ and $B$, 
i.~e.~Eq.~(\ref{eq-1}) also holds with $\pt\rho$. If we can
``switch'' the partial transpose sign from $\rho$ to $A$ and $B$ in
this $\pt\rho$ uncertainty relation, we obtain the
Schr\"odinger-Robertson partial transpose (SRPT) inequality
\begin{eqnarray}  (\Delta \pt A)^2 (\Delta \pt B)^2 \ge \frac{1}{4} | \langle [A,B\pt ] \rangle |^2 \nonumber
\\  \ +  \frac{1}{4} | \langle \{ A,B \pt \} \rangle  - 2 \langle \pt A \rangle \langle \pt B \rangle |^2 , \label{eq-2}
\end{eqnarray}
that is never violated for separable states and a violation of this
inequality is a sufficient condition to entanglement.  The key result of this paper is that
Eq.~(\ref{eq-2}) has the property of being
experimentally implementable since it deals with observable
quantities, as the partial transpose of any observable remains an
observable.

However, ``switching'' the partial transpose is not a trivial
operation since it must take place in two types of terms~: we need
our observable $A$ to hold $\tr{\pt \rho A}= \tr{ \rho \pt A}$ and
$\tr{\pt \rho A^2}=\tr{ \rho (\pt A)^2 }$ for any state $\rho$ (and
similarly for $B$). It turns out the first condition always holds
and the second is only verified for specific observables.
Eventually, we will be able to prove the following result:

{\bf Proposition 1.} For any entangled bipartite pure state
$\ket\psi \in \mathcal{H}_1\otimes \mathcal{H}_2$, with
$\mathcal{H}_1$ and $\mathcal{H}_2$ two Hilbert spaces of any
dimension, there are observables $A$, $B$ acting on $\mathcal{H}_1\otimes \mathcal{H}_2$ satisfying
\[ (\pt {A})^2 =\pt{ \left(A^2 \right)}, \; (\pt {B})^2 = \pt{\left(B^2 \right)}, \label{eq-3}\]
such that the SRPT inequality (\ref{eq-2}) is violated.

Before proving this result, we need to show that conditions (\ref{eq-3}) are necessary for Eq. (\ref{eq-2}) to be valid. The next two lemmas do so explicitly.

{\bf Lemma 1.} The relation $ \tr{\pt \rho A} = \tr{ \rho \pt A}$
holds for any density operator $\rho$ and any observable $A$.

\emph{Proof.} Using matrix components defined as
$ M_{i \mu,j \nu} \equiv \langle i,\mu | M | j,\nu \rangle$, with $\{\ket i\}$ the basis of the first subsystem and $\{\ket \mu\}$ the basis of the rest of the system, and noting that the partial transposition effect is to swap the latin indices, i.e., $(\pt M)_{i\mu,j\nu} = M_{j\mu,i\nu}$, we have
$ \tr{\pt \rho A} =  \rho_{l\mu,i\lambda} A_{l \lambda,i \mu} = \tr{ \rho \pt A},$
where there is a sum on repeated indices (notation adopted
throughout the paper).

{\bf Lemma 2.} The relation $ \tr{\pt \rho A^2} = \tr{ \rho (\pt A)^2 },$
holds for any density operator $\rho$ if and only if the observable $A$
obeys $ (\pt {A})^2 = \pt{\left(A^2\right)} .$

\emph{Proof.} The ``if'' part is a direct consequence of lemma 1.
The ``only if'' part is obtained as follows~:
\begin{eqnarray}
\tr{\pt \rho A^2}&=& \ \rho_{l\mu,i\lambda} A_{l\lambda,k\sigma} A_{k\sigma,i\mu} \\
\tr{ \rho (\pt A)^2 }&=& \ \rho_{l\mu,i\lambda}  A_{k\lambda,i\sigma} A_{l\sigma,k\mu}.
\end{eqnarray}
If those quantities are equal for any $\rho$, i.e. if the quantity
$ \rho_{l\mu,i\lambda} ( A_{l\lambda,k\sigma} A_{k\sigma,i\mu} -  A_{k\lambda,i\sigma} A_{l\sigma,k\mu} )$
is zero for any $\rho_{l\mu,i\lambda}$, then the term between parenthesis
must be zero and we conclude.

In general, observables do not  satisfy Eq.~(\ref{eq-3}), which can result in a violation of an
SRPT inequality applied on a separable state with unsuitable observables. To illustrate this, we consider
the inequality corresponding to the computational basis vector
$\ket{00}$ of a two qubit system using the observables $A =
\sigma_x\otimes\sigma_x$, and $B = \sigma_x
\otimes \sigma_y +  \sigma_y \otimes \sigma_x $, with usual notations, which is
violated, since $(\pt {B})^2 \neq \pt{(B^2)}$.
The former example illustrates the importance of using suitable
observables in the SRPT inequality. We now prove proposition 1.

\emph{Proof of proposition 1.} Let us consider an entangled state $\ket\psi$ and express it in the following decomposition:
\[ \ket\psi=\sum_i c_i \ket{i}_1\otimes \ket{i}_2,\]
where the $\ket i_j$ are a basis of $\mathcal{H}_j$ and the $c_i$ complex numbers. Such a decomposition is always possible, the Schmidt decomposition being a particular one with real $c_i$ coefficients.
We will work in the $\ket{ij} \equiv \ket{i}_1\otimes \ket{j}_2$ basis, expressing operators through that basis.

 Since $\ket\psi$ is entangled, there are at least two non-zero coefficients; let us assume without loss of generality $c_0 \neq 0 \neq c_1$. We define two observables
\[  A=\ketbra{01}{01}, \;  B=\sigma_x\otimes\sigma_x,\label{eq-11}\]
 where $\sigma_x \equiv \ketbra{0}{1} + \ketbra{1}{0}$. $A$ and $B$ can be checked to obey Eq.~(\ref{eq-3}) and using state $\ket\psi$ we further find
 \begin{eqnarray}
(\Delta \pt A)^2  &=& 0 \\
\frac 1 4 | \langle \pt{[ A,B]} \rangle |^2 &=& \mbox{Im}(c_0^* c_1)^2 \\
\frac 1 4 | \langle \pt{\{A,B\}} \rangle  - 2 \langle \pt A \rangle \langle \pt B \rangle |^2 &=& \mbox{Re}(c_0^* c_1)^2.
\end{eqnarray}

The SRPT inequality is then written
\[ 0 \ge  \mbox{Re}(c_0^* c_1)^2 + \mbox{Im}(c_0^* c_1)^2 =  |c_0|^2 |c_1|^2,  \label{eq-9} \]
and is always violated for non-zero $c_0$ and $c_1$. We will discuss the case of mixed states in the latter part of this paper.

In two-qubits systems, the general form of observables satisfying Eq.~(\ref{eq-3}) is extremely simple~:

{\bf Proposition 2.} For any $2 \times 2$ observable $M$, the relation
\[(\pt {M})^2 = \pt{\left(M^2\right)}, \label{eq-7}\]
holds if and only if it can be written as
\[ M = (\mathbf{a}\cdot\boldsymbol{\sigma})  \otimes  (\mathbf{b}\cdot\boldsymbol{\sigma})  + \id \otimes (\mathbf{c}\cdot\boldsymbol{\sigma}) + (\mathbf{d}\cdot\boldsymbol{\sigma}+\eta \, \id ) \otimes \id ,\]
where $\boldsymbol{\sigma}$ is the vector composed of the 3 Pauli
operators, $\id$ is the identity operator, $\mathbf{a}, \mathbf{b}, \mathbf{c}$ and $\mathbf{d}$ are
four real vectors and $\eta$ is a real number.

\emph{Proof.} Let us write the observable $M$ as
$M = a_{\mu\nu} \, \sigma_\mu \otimes \sigma_\nu,$
where the $\sigma_\mu$ ($\mu \in \{0,1,2,3\}$) are the Pauli operators, $\sigma_0 \equiv \id$ and the $a_{\mu\nu}$ are real coefficients defined as~$ a_{\mu\nu} = \frac 1 4 \tr{M \, \sigma_\mu \otimes \sigma_\nu}$. If we express the condition $(\pt {M})^2 - \pt{(M^2)}=0$ in that basis, we find that there are no constraint on the $a_{\mu\nu} $ coefficient with a 0 index ($\mathbf{c},\mathbf{d}$ and $\eta$ take any values) and we are left with the condition $\epsilon_{ijn}\epsilon_{klm} a_{ik}  a_{jl}=0$, for all $m,n$, which  expresses that every $2\times 2$ minor of the $3\times 3$ matrix $a_{ij}$ must be zero. Therefore all the lines (or columns) of that matrix must be linearly dependent and we can write $ a_{ij}=a_i b_j$ (with $a_i$ and  $b_j$ the components of $\mathbf{a}$ and $\mathbf{b}$).

 As a particular case of this proposition, we note that if $M$ can be written as $M_1 \otimes M_2$ with $M_1$ and $M_2$ two observables, then $M$ satisfies Eq.~(\ref{eq-7}); furthermore, it is straightforward to show that, in this case, $M_1$ and $M_2$ can be of any dimension and $M_2$ can even act on more than one subsystem.

 We will now discuss some applications of the SRPT inequality in the bipartite case.

{\bf 2D harmonic oscillator.} We consider entanglement in states of a two dimensional oscillator with definite energy and angular momentum (see e.~g.~Ref.~\cite{Mai01} describing the experimental production of entangled angular momentum states of photons). Those states are the common eigenvectors $\ket{\psi_{k,M}}$ ($k, M$ integers) of the hamiltonian $H=\omega(a a^\dagger + b b^\dagger +1)$ ($a$ and $b$ are the oscillator annihilation operators) and the angular momentum $L_z = i (a b^\dagger - a^\dagger b)$ with eigenvalues $\omega(n+1)$ (with $n=2k + |M|$) and $M$, respectively .
The states $\ket{\psi_{k,M}}$ can always be expressed in the number state basis $\ket{n_1,n_2}$ as $\sum_{i=0}^n c_i \ket{i,n-i}$ with non-zero $c_0$ and $c_n$ coefficients. They are thus clearly entangled for $n > 0$. This entanglement is well detected by the pair of observables
\[  A=\ketbra{00}{00}, \;  B=\sigma^{(n)}_x\otimes\sigma^{(n)}_x, \label{eq-27}\]
with $\sigma_x^{(n)} \equiv \ketbra{0}{n} +\ketbra{n}{0}$, which yields for those states the SRPT inequality $|c_0| |c_n| \le 0$, evidently violated.

{\bf Multiphoton polarization state.}
For some particular experiments, the SRPT inequality can be particularly efficient. Here, we show that on some multiphoton polarization states, the detection of entanglement only involves the measurement of two projectors. Let us consider the entangled two-photon state
\[ \ket\psi = \alpha \ket{0,2} + \beta \ket{1,1} + \gamma \ket{2,0}, \label{eq-multi} \]
where $\alpha, \beta, \gamma$ are arbitrary coefficients such that $ \mbox{Re}(\alpha^* \gamma)\neq 0$ and $\ket{m,n}$ denotes $m$ photons in a given polarization state and $n$ photons orthogonally polarized to the $m$ firsts. The production and properties of those states have been studied in~\cite{Tse00}. Using the observables
\[  A=\ketbra{00}{00}, \;  B= \sigma^{(2)}_x\otimes\sigma^{(2)}_x, \]
and dropping the commutator term in (\ref{eq-2}), we get the inequality $0 \ge | \mbox{Re}(\alpha^* \gamma)|$. Since $\ket\psi$ is never the vacuum, we have $\langle \pt A \rangle = \Delta \pt A = 0$ and $\pt B$ does not need to be measured. All that is needed to detect entanglement is the measurement of  $\{A,B\pt \}= \ketbra{\psi^+}{\psi^+}-\ketbra{\psi^-}{\psi^-}$ with $\ket{\psi^\pm}\equiv (\ket{02} \pm \ket{20})/\sqrt 2$. More generally, the entanglement of an $N$-photon state of the form $\sum_{i=0}^N c_i \ket{i,N-i }$  will always be easily detectable with similar observables.

{\bf Cat states.} We consider the normalized Schr\"odinger cat state 
\[ \ket{\psi}=(\ket{\alpha, \beta} + \ket{-\alpha,-\beta})/\mathcal{N},\]
 where $\ket\alpha, \ket\beta$ are coherent states and $\mathcal{N}=\sqrt{2+2 \exp(-2|\alpha|^2-2|\beta|^2)}$. The state $\ket{\psi}$ is a bipartite even state whose production and properties are discussed in~\cite{Ger07}. Experimentally, it is possible to show the entanglement of $\ket\psi$ with the quadrature operators
\begin{eqnarray}
A&=& a_1 (a^\dagger +a) + b_1 (b^\dagger +b), \\
 B&=& i a_2 (a^\dagger - a) + i b_2 (b^\dagger -b),
 \end{eqnarray}
where $a$ and $b$ are the annihilation operators of the fields and $a_i$ and $b_i$ are real parameters. We can assume $\alpha$ and $\beta$ to be real, then we get $(\Delta \pt A)^2= a_1^2+b_1^2 + 8 (a_1 \alpha + b_1 \beta)^2/ \mathcal{N}^2$, $(\Delta \pt B)^2= a_2^2+b_2^2 - 4  (a_2 \alpha - b_2 \beta)^2/ (1+\exp(2\alpha^2+2\beta^2))$, $ \frac{1}{4} | \langle [ A,B \pt ] \rangle |^2= (a_1 a_2 + b_1 b_2)^2 $ and the anticommutator term is zero. Setting $a_1 = - a_2 = - \beta$ and $b_1=-b_2= \alpha$ insures a violation of the SRPT inequality for non-zero $\alpha$ and $\beta$.

In order to compare the results, one may try to apply the entanglement criterion introduced by Duan \textit{et al.} ~\cite{contin} on $\ket{\psi}$. That criterion is a sufficient condition for entanglement and is also necessary when applied on gaussian states. Clearly, the state $\ket\psi$ is not gaussian, but the criterion may still be applied. The calculation is very close to the previous on, however it can be shown that the cat state $\ket\psi$ never violates Duan \textit{et al.}'s inequality.

The SRPT inequality is also a strong criterion in the tripartite case. We recall that a tripartite pure state is fully separable if it can be written as a combination of three separable subsystems, biseparable if it can be written as a combination of one subsystem separated from the other (entangled) subsystems and fully entangled otherwise. 
In that last case, for three qubits, there are two separate classes of entanglement represented by the states $\ket{\mbox{GHZ}}$ and $\ket{\mbox{W}}$~\cite{Dur00}.

Any three-qubit state can always be written under the form~\cite{Aci00}:

\[\ket{\psi} = \lambda_0 \ket{000}+\lambda_1 \ket{100}+\lambda_2 \ket{101}+\lambda_3 \ket{110}+\lambda_4 \ket{111}, \label{eq-3qb} \]
where one $\lambda_i$ is complex and the other ones are real. When  all $\lambda_i=0$ but $\lambda_0$ and $\lambda_4$, the state is of the GHZ-type. We get to our next result:

{\bf Proposition 3.} For any entangled tripartite pure state, there are observables satisfying Eq.~(\ref{eq-3}) such that a Schr\"odinger-Robertson  partial transpose inequality is violated.

\emph{Proof.} We first consider a three-qubit state and express it as in Eq.~(\ref{eq-3qb}). 
The three sets of observables
\begin{eqnarray}
A=\ketbra{001}{001} &\;&  B=\sigma_x\otimes\id\otimes\sigma_x,\\
A=\ketbra{010}{010} &\;&  B=\sigma_x\otimes\sigma_x\otimes\id,\\
A=\ketbra{011}{011} &\;&  B=\sigma_x\otimes\sigma_x\otimes\sigma_x,
\end{eqnarray}
lead to the SRPT inequalities $|\lambda_0||\lambda_2| \leq 0$, $|\lambda_0||\lambda_3| \leq 0$ and $|\lambda_0||\lambda_4| \leq 0$. If $\lambda_0 =0$ the inequalities are not violated, but in that case $\ket\psi = \ket{1} \otimes (\lambda_1 \ket{00}+\lambda_2 \ket{01}+\lambda_3 \ket{10}+\lambda_4 \ket{11})$ and is biseparable. We already know that every entangled two-qubit state can be detected with the mean of an SRPT inequality. If $\lambda_2=\lambda_3=\lambda_4=0$, there is no violation of the inequalities either, but in that case $\ket\psi = (\lambda_0 \ket{0} + \lambda_1 \ket{1}) \otimes \ket{00}$ and is fully separable. Therefore, there is always an SRPT inequality able to detect the entanglement of $\ket\psi$. 
In a Hilbert space of dimension greater than $2\times 2 \times 2$, a straightforward generalization of the demonstration in~\cite{Aci00} shows that any state $\ket\varphi$ can always be written as $\ket\psi + \ket{\psi'}$ with $\ket{\psi'}$ a linear combination of all basis product bases $\ket{n_1 n_2 n_3}$ with at least one $n_i > 1$. In that case, the observables we gave ignore $\ket{\psi'}$ and the result holds.

An interesting result is the fact that a pair of bipartite operators can never detect a GHZ-type state. Indeed, the expectation value of an observable $A$ on a GHZ-type state expressed as in Eq.~(\ref{eq-3qb}) will be a combination of the terms $\langle 000 |A| 000\rangle$, $\langle 000 |A| 111\rangle$, $\langle 111 |A| 000\rangle$ and $\langle 111 |A| 111\rangle$. If $A$ is a bipartite observable acting, e.~g., on the first two subsystems, we have $\langle 000 |A| 111\rangle = \langle 00 |A_{12}| 11\rangle  \langle 0|\id|1\rangle = 0$. Thus, the mean value of the observable $A$ acting on a GHZ-type state is the same as if $A$ were acting on a separable state of the form $\rho=\lambda_0^2 \ketbra{000}{000} + \lambda_4^2 \ketbra{111}{111}$. Therefore there cannot be any violation of an SRPT inequality.

Next we discuss some applications on tripartite and multipartite systems.

{\bf 3D harmonic oscillator.} We consider entanglement in the angular momentum states of a three-dimensional harmonic oscillator. Those states are the common eigenvectors $\ket{\psi_{k,l,m}}$ ($k, l, m$ integers, $|m| \le l$) of the hamiltonian $H=\omega(a a^\dagger + b b^\dagger + c c^\dagger + 3/2)$
($a$, $b$, $c$ are the oscillator annihilation operators according to the 3 directions $x$, $y$ and $z$, respectively), the squared total angular momentum $L^2$~\footnote{$L^2=-(a^2 b^{\dagger 2} +a^{\dagger 2} b^2 +a^2 c^{\dagger 2} + a^{\dagger 2} c^2+b^2 c^{\dagger 2} + b^{\dagger 2} c^2) + 2(a a^\dagger b b^\dagger +a a^\dagger c c^\dagger + b b^\dagger c c^\dagger+ a a^\dagger + b b^\dagger + c c^\dagger)$.}, and the angular momentum $z$-component $L_z =i (a b^\dagger - a^\dagger b)$ with eigenvalues $\omega(n + 3/2)$ (with $n=2k + l$), $l(l+1)$ and $m$, respectively. The states $\ket{\psi_{k,l,m}}$ can always be expressed in the number state basis $\ket{n_1,n_2,n_3}$ as $\sum_{i=0}^n\sum_{j=0}^i c_{ij} \ket{j,i-j,n-i}$ and
are entangled for $\ket{\psi_{k,l,m}}=\ket{\psi_{0,1,\pm 1}}$ or whenever $n > 1$. In this case,
the $m=0$ (resp.~$m\neq 0$) states are characterized with non-zero coefficients $c_{20}, c_{22}$ (resp.~$c_{m0}, c_{mm}$).
The entanglement of the $m=0$ states can be detected using the observables
\[ A=\ketbra{00n-2}{00n-2},  B=\sigma^{(2)}_x\otimes\sigma^{(2)}_x\otimes \ketbra{n-2}{n-2}, \]
which yield the SRPT inequality $|c_{20}||c_{22}|\le 0$, evidently violated. For $m\neq 0$, the observables
\[ A=\ketbra{00n-m}{00n-m}, B=\sigma^{(m)}_x\otimes\sigma^{(m)}_x\otimes \ketbra{n-m}{n-m}. \]
yield similarly the violated SRPT inequality $|c_{m0}||c_{mm}|\le 0$.

{\bf Bipartite Werner states.}  The generalization of proposition 1 to mixed states is a
difficult task. We may try, as an illustrative
example, to detect the Werner mixed state $ \rho_W = x \ketbra{\psi}{\psi} + \id (1-x)/4$~\cite{Wer89}
with the normalized state $\ket{\psi} = a \ket{00} + b \ket{11}$.
The PPT criterion shows that $ \rho_W$ is entangled if and only if
$ x > 1 / (1+4|a||b|).$ Using the observables:
\[  A=\sigma_z\otimes\sigma_z, \;  B=\sigma_x\otimes(\cos\varphi\, \sigma_x+\sin\varphi \, \sigma_y),\]
our SRPT inequality detects the entanglement of  $\rho_W$ when
$x > 2/(1+\sqrt{1+32\, \mbox{Re}(e^{i \varphi} a^* b)})$.

In the particular case when $\ket\psi$ is the Bell state $\ket{\phi^\pm}$ and $\varphi=0$, $\rho_W$ is entangled iff $x > 1/3$ whereas it is detected via the SRPT inequality  when $x > 1/2$. This result improves the limits of detection given by the Bell inequalities ($x > 1 / {\sqrt 2}$, see~\cite{Per96}) or by the uncertainty relations of G\"uhne~\cite{Guh04} ($x> 1/{\sqrt 3}$).

{\bf  Multipartite Werner states.} The SRPT inequality can be applied on mixed states of multipartite systems. Let us look at its results on the $N$-dimensional Werner mixed state $\rho(x)= x \ketbra{\mbox{GHZ}_N}{\mbox{GHZ}_N} + \id (1-x)/2^N$, with $\ket{\mbox{GHZ}_N} \equiv (\ket{0\cdots0}+\ket{1\cdots1})/\sqrt 2$.  Using the observables
\begin{eqnarray}
A&=&\ketbra{01\cdots 1}{01\cdots1}+\ketbra{10\cdots 0}{10\cdots 0},\\
B&=&\ketbra{0\cdots 0}{1\cdots1}+\ketbra{1\cdots 1}{0\cdots 0} \nonumber \\
 && \ +\ketbra{01\cdots 1}{10\cdots0}+\ketbra{10\cdots 0}{01\cdots 1},
\end{eqnarray}
we find an SRPT inequality violated if $x> 1/(1+2^{N-2})$. The PPT criterion gives the sufficient limit of entanglement $x> 1/(1+2^{N-1})$ and we can find in~\cite{Tot05} a witness giving the detection limit  of $x>(2-2^{2-N})/(3-2^{2-N})$, which is strictly smaller than our result for $N \geq 3$ (also, that limit approaches $2/3$ as $N$ grows, where ours approaches 0). For $N=3$, the PPT criterion gives $x>1/5$, we find the limit $x>1/3$ while in~\cite{Tot05} the limit is $x>3/5$ and another witness in~\cite{Guh04} gives the limit $x>3/7$.

Finally it should be kept in mind that any criterion based on inequalities would be restrictive as these are based on two chosen observables unlike the density operators which contain all the information. One could of course increase the number of observables and work a stronger criterion
 based on the positivity condition $\langle \left( \sum_i c_i A_{i} \right)^\dagger
  \left(\sum_i c_i A_{i} \right)  \rangle \ge 0$~\cite{Ush07},
for any observable $A_{i}$.
Further possibilities consist of using generalized uncertainty relations
which are especially suitable for mixed states~\cite{Aga03}. 

In conclusion, we have developed in this paper an entanglement
criterion based on a Schr\"odinger-Robertson type inequality mixed
with positive partial transpose operations (the SRPT inequality). In
this context, we have proven that all observables are not suitable
for entanglement detection and we have given the general condition
they must satisfy. For $2 \times 2$ bipartite systems, we have explicitly
given their structure. We have shown that, for any pure bipartite and tripartite state, we
can always find observables leading to a violation of the SRPT
inequality. We have applied the inequality on a variety of physical systems, including
bipartite Werner states, for which we found a better detection limit than
Bell type inequalities, and multipartite Werner states, for which we also found a good detection limit.

\acknowledgments{J.~G. thanks the Belgian F.R.S.-FNRS for financial
support and thanks G.~S.~A.~for the hospitality at the Oklahoma State University.}
	

\end{document}